\documentclass[conference]{IEEEtran}
\pdfoutput=1    

\usepackage{xcolor}
\definecolor{wong-black}        {HTML}{000000}
\definecolor{wong-lightorange}  {HTML}{E69F00}
\definecolor{wong-lightblue}    {HTML}{56B4E9}
\definecolor{wong-green}        {HTML}{009E73}
\definecolor{wong-yellow}       {HTML}{F0E442}
\definecolor{wong-darkblue}     {HTML}{0072B2}
\definecolor{wong-darkorange}   {HTML}{D55E00}
\definecolor{wong-pink}         {HTML}{CC79A7}

\usepackage{booktabs}
\usepackage[accsupp]{axessibility}  

\usepackage{url}

\usepackage{hyperref} 
\hypersetup{
    colorlinks=true,
    citecolor=wong-green,
    linkcolor=wong-darkblue,
    filecolor=wong-pink,      
    urlcolor=wong-black,
    pdfpagemode=FullScreen,
    }

\newcommand{\figref}[1]{Fig.~\ref{#1}}

\usepackage{array}
\newenvironment{conditions}
  {\par\vspace{\abovedisplayskip}\noindent\begin{tabular}{>{$}l<{$} @{${}={}$} l}}
  {\end{tabular}\par\vspace{\belowdisplayskip}}

\usepackage{cite}
\usepackage{amsmath,amssymb,amsfonts}
\usepackage{algorithmic}
\usepackage{graphicx}
\usepackage{textcomp}
\usepackage[nolist, nohyperlinks, printonlyused]{acronym} 
\usepackage{pifont}
\newcommand{\xmark}{\ding{55}}

\def\BibTeX{{\rm B\kern-.05em{\sc i\kern-.025em b}\kern-.08em
    T\kern-.1667em\lower.7ex\hbox{E}\kern-.125emX}}
    
\newcommand\nnfootnote[1]{  
  \begin{NoHyper}
  \renewcommand\thefootnote{}\footnote{#1}%
  \addtocounter{footnote}{-1}%
  \end{NoHyper}
}

\usepackage{tikz}
\usetikzlibrary{arrows.meta,
                positioning,
                shapes}
\usepackage[normalem]{ulem}
\usepackage{listings}

\begin{document}


\title{{Impact, Attention, Influence: \\Early Assessment of Autonomous Driving Datasets}}



\author{\IEEEauthorblockN{Daniel Bogdoll\IEEEauthorrefmark{2}\IEEEauthorrefmark{3}\textsuperscript{\textasteriskcentered},
    Jonas Hendl\IEEEauthorrefmark{3}\textsuperscript{\textasteriskcentered},
    Felix Schreyer\IEEEauthorrefmark{2},
    Nishanth Gowda\IEEEauthorrefmark{2},
    Michael Färber\IEEEauthorrefmark{3}
    and J. Marius Zöllner\IEEEauthorrefmark{2}\IEEEauthorrefmark{3}}

  \IEEEauthorblockA{\IEEEauthorrefmark{2}FZI Research Center for Information Technology, Germany\\
    bogdoll@fzi.de}
  \IEEEauthorblockA{\IEEEauthorrefmark{3}Karlsruhe Institute of Technology, Germany}}

\maketitle

\nnfootnote{\textasteriskcentered~These authors contributed equally}


\begin{acronym}
  \acro{ad}[AD]{Autonomous Driving}
  \acro{jif}[JIF]{journal impact factor}
  \acro{IS}{Influence Score}
  \acro{gbif}[GBIF]{Global Biodiversity Information Facility}
  \acro{fair}[FAIR]{Findable, Accessible, Interoperable, Reusable}
\end{acronym}


\begin{abstract}
  \ac{ad}, the area of robotics with the greatest potential impact on society, has gained a lot of momentum in the last decade. As a result of this, the number of datasets in \ac{ad} has increased rapidly. Creators and users of datasets can benefit from a better understanding of developments in the field. While scientometric analysis has been conducted in other fields, it rarely revolves around datasets. Thus, the impact, attention, and influence of datasets on autonomous driving remains a rarely investigated field. In this work, we provide a scientometric analysis for over 200 datasets in \ac{ad}. We perform a rigorous evaluation of relations between available metadata and citation counts based on linear regression. Subsequently, we propose an Influence Score to assess a dataset already early on without the need for a track-record of citations, which is only available with a certain delay.


\end{abstract}




\section{Introduction}
\label{sec:introduction}

Autonomous driving technology does not only affect urban transportation~\cite{waymo_2022} and delivery of goods~\cite{waabi_2022}, but also farming~\cite{farming_2022} or warehouse logistics~\cite{warehouse_2022}. With the progress of deep learning and this growing interest in \ac{ad} in many fields of robotic, the number of related datasets is consistently increasing. The datasets have also increased in size and many have become increasingly specialized~\cite{Bogdoll_Addatasets_2022_VEHITS}. The most extensive collection of datasets known to us, \textit{ad-datasets}, currently lists 231 datasets in the domain~\cite{addatasets_web}. However, not all of them are being equally used in the robotics community, the distribution of their citations is heavily skewed. As part of the more impactful works, well known datasets for the core tasks perception and prediction dominate~\cite{geiger_vision_2013,sun_scalability_2020, qiao_vip-deeplab_2021}. As part of the long tail, many datasets for niche research areas exist~\cite{maeda_road_2018, codevilla_exploring_2019, lehner_3d-vfield_2022}. Well known datasets tend to bring many advantages with them: They enable comparison between works, have higher quality, advanced tooling, and often community knowledge and support is available. The increasing number of datasets, which are potentially interesting but lack reputation, leads to a lot of untapped potential: Many researchers are hesitant to use such datasets and stick to old, but established ones instead~\cite{addatasets_web}. This is why we asked ourselves the question: Given a novel dataset without a multi-year track record of citations, is there a way to estimate its future development? Datasets with a high potential might be more appealing already in their early days.

\begin{figure}
    \includegraphics[width=\columnwidth]{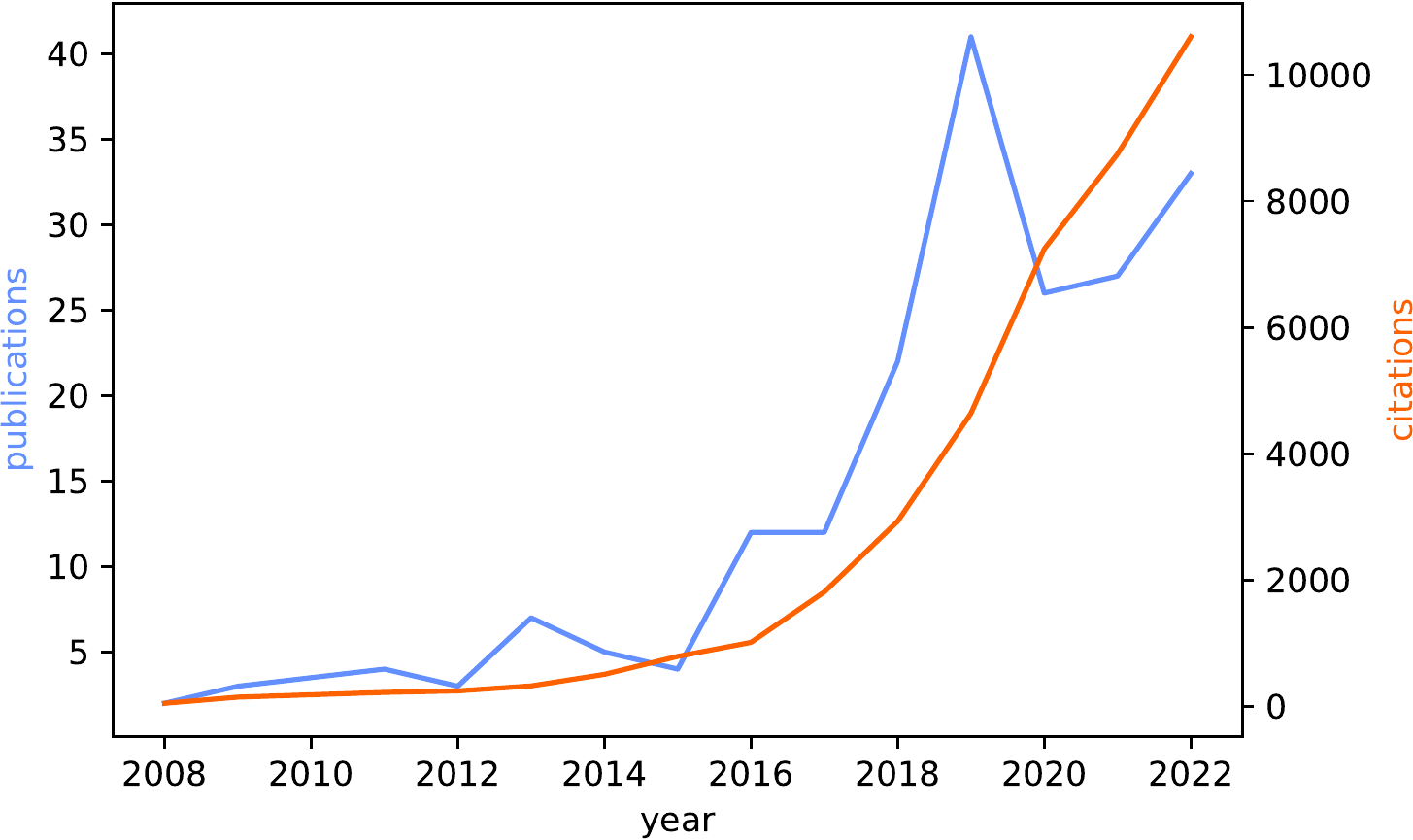}
    \centering
    \caption{Course of published datasets and citations of the accompanying publications in the domain of \ac{ad}. This growing number of datasets, initially without reputation, holds a great deal of untapped potential as researchers struggle to use new datasets for their research. Datasets as listed on ad-datasets~\cite{addatasets_web}, citation counts from Semantic Scholar~\cite{allen_institute_for_ai_semantic_nodate}.
    }
    \label{fig:growth_of_publications}
\end{figure}

\textbf{Research Gap.} To date, citations are mostly used to assess datasets, which are not available early on. Thus, new datasets can have a hard time gaining traction, which results in untapped potential. It is yet not well understood if and how metadata of datasets relate to future impact or how they can be utilized to assess datasets early on. To the best of our knowledge, such an analysis has not yet been performed.

\textbf{Contribution.}
In order to analyze the field of datasets, we first assembled the largest collection of datasets with enriched metadata available, including over 200 datasets with metadata from three different sources. We then applied linear regression to evaluate factors which relate to the future impact of datasets, measured in citations. Finally, we propose the \ac{IS}, which is a mean to assess datasets early on without the need of a multi-year track record of citations. The \ac{IS} can be used to assess any dataset at any given year, which also allows for later analysis. Our work aims to help researchers from the robotics community to better understand and assess the performance of datasets. This can lead to the design of better and thus more influential datasets as well as an actionable analysis of new datasets to assess their potential. All data used in this work is as of January 04, 2023. All code is available on 
\href{https://github.com/daniel-bogdoll/ad_datasets_influence}{\color{wong-lightblue}{GitHub}}.

\section{Related Work}
\label{sec:related_work}

Here, we give an introduction to scientometrics, bibliometrics, and altmetrics, followed by dataset analysis.

\subsection{Scientometrics, Bibliometrics, and Altmetrics}
Scientometrics, Bibliometrics, and Altmetrics are highly intertwined fields that focus on the analysis of science and its processes as a whole, written works of science, and online communication of science, respectively~\cite{chellappandi_bibliometrics_2018}.

\textbf{Scientometrics.}
Ravenscroft et al.~\cite{ravenscroft_measuring_2017} examined the impact of research by comparing citation-based metrics, such as citation count or h-index~\cite{hirsch_index_2005}, with altmetrics and impact other than citations, e.g., societal and economic impact. However, they found no strong relationship between the fields. Hicks et al.~\cite{hicks_bibliometrics_2015} suggest using multiple factors to portray multiple aspects.

Leydesdorff et al.~\cite{leydesdorff_citations_2016} claim that citations are equated to impact and evaluate the relationship between impact and research quality. They found that short-term citations signify the investment in a current discourse, while long-term citations signify acceptance as reliable scientific knowledge.  However, some researchers question if or to what extent citations measure scientific impact and point to issues, e.g., inconsistent reasons for citations~\cite{bornmann_what_2008}. Problems include the cumulative advantages already successful papers experience~\cite{price_general_1976}, self-citations, which men do more often~\cite{chawla_men_2016}, negative citations, and citing out of reasons that do not reflect actual use or relevance~\cite{mingers_review_2015}. Valenzuela et al.~\cite{Valenzuela2015IdentifyingMC} presented a method to identify four types of citations: "Related work, Comparison, Using the work, Extending the work"~\cite{Valenzuela2015IdentifyingMC}, which is used by Semantic Scholar to determine “Highly Influential Citations”~\cite{semanticscholar_highly}. However, it shows a high correlation with citations. 

The field of trend detection analyzes large corpora of works to detect upcoming patterns~\cite{leDetectingEmergingTrends2005, salatinoEarlyDetectionForecasting, farberFindingTemporalTrends, farberScholarSightVisualizingTemporal2019}. Lopez Belmonte et al.~\cite{lopez_belmonte_machine_2020} analyzed publications in Machine Learning and Big Data and found exponential growth of publications. They compared the popularity of keywords and the h-index.

\textbf{Bibliometrics.}
Citations can be aggregated on different levels, e.g., for the papers of one author as the h-index does, or on the journal level, like the \ac{jif}, which is the two-year average ratio of citations to articles published. The \ac{jif} is ill-suited for evaluating individual papers by means of the journal it was published in~\cite{paulus_impact_2018}. This is due to the heavy skewness of the distribution of citation counts within journals~\cite{callaway_beat_2016}. The Hirsch-index, usually referred to as the h-index, combines the productivity of an author with the impact of their individual papers. Using the h-index increases robustness compared to simply counting the total number of citations, as few highly cited papers have little effect on the h-index.
In addition, there have been efforts to recommend papers and citations~\cite{Beel2016ResearchpaperRS, Frber2018ToCO}, predict future citation counts of papers~\cite{fu_using_2010,pobiedina_citation_2016, ma_deep-learning_2021, li_deep_2019} and the impact of scientists~\cite{butun_predicting_2020}. Such approaches remain challenging and are often domain-specific.

Bornmann and Marx~\cite{bornmann_proposal_2013} have proposed to expand the bibliometric analysis by not only considering citations but also references. Following this idea, reference analysis has been used to identify influential references~\cite{yeung_monoamine_2019}.

\textbf{Altmetrics.}
Online interactions with papers are referred to as altmetrics and are usually available earlier than citations, which gives altmetrics an advantage over bibliometrics~\cite{bornmann_altmetrics_2014}. Bornmann and Marx~\cite{bornmann_altmetrics_2018} examined if Altmetrics can be used to predict paper quality which was measured through peer assessments and found that both tweets and readers do, with the latter having a stronger relationship. Lamb et al.~\cite{Lamb_tweet_2018} showed that the Altmetric Attention Score is a predictor of the citations of a paper in ecology and conservation. 
 Zavrel~et~al.~\cite{zavrel_can_2022} clustered papers released at the International Conference on Machine Learning (ICML) in 2022 and calculated a score for their impact. They used Twitter mentions, citations, the authors' average h-index, and an award for outstanding papers rewarded by the conference itself. They claim to do a “simple combination of these four scores to calculate an impact score”~\cite{zavrel_can_2022} but do not reveal the formula. Färber analyzed GitHub repositories of papers, mostly from the field of AI, and found a power-law distribution of stars and forks~\cite{10.1145/3383583.3398578}. While Haustein et al. claim that "Altmetrics measures scientific impact based on online references and activity"~\cite{haustein_coverage_2014}, many disagree with equating altmetrics with impact. For example, Sugimoto states that "attention is not impact" and calls online interaction with scientific works "attention"~\cite{sugimoto_attention_2015}. Altmetrics might reflect broader or societal impact~\cite{bornmann_altmetrics_2014}.

\subsection{Dataset Analysis}
Bogdoll et al.~\cite{Bogdoll_Addatasets_2022_VEHITS} gathered metadata of over 200 datasets in the field of autonomous driving. Similarly, Färber and Lamprecht released the data set knowledge graph, which is a collection of over 2,000 datasets with added metadata~\cite{farberDataSetKnowledge2021}. D'Ulizia et al.~\cite{dulizia_fake_2021} analyzed the metadata of datasets for fake news detection. Utamachant and Anutariya~\cite{utamachant_analysis_2018} analyzed the datasets of Thailand's national open data portal, but relied on domain experts to assess impact. Nguyen and Weller proposed FAIRnets, a service to search for neural networks and their related datasets~\cite{nguyen2019fairnets} published on GitHub. They build upon the \ac{fair} principles~\cite{Wilkinson2016}, which "put specific emphasis on enhancing the ability of machines to automatically find and use the data"~\cite{Wilkinson2016}. 
Khan et al.~\cite{khan_measuring_2021} analyzed datasets from the \ac{gbif} which publishes datasets with a DOI and indexes datasets in biodiversity. They promote data standards and the reuse of datasets~\cite{global_biodiversity_information_facility_what_nodate} as well as accompanying publications, which they call "data papers", that describe a dataset thoroughly~\cite{global_biodiversity_information_facility_data_nodate}.
Khan et al.~\cite{khan_measuring_2021} report a strong correlation between dataset download numbers and citation counts, and suggest that downloads and citations signify a similar kind of impact. They also find correlations between altmetrics and citations. Moreover, they question whether every citation means the usage of a dataset and point to differences in citation behavior.
F\"arber et al. proposed an approach to find methods and datasets which authors actually used when citing the related paper~\cite{farber_identifying_2021}. However, unrealistically few dataset usages were identified.

\tikzstyle{data} = [cylinder, minimum width=2cm, minimum height=1cm, text centered, draw=black, fill=white, text width=3cm, shape border rotate=90, aspect=0.1]
\tikzstyle{process} = [block, minimum width=4cm, minimum height=1.5cm, text centered, draw=black, fill=white, text width=3cm]
\tikzstyle{io} = [trapezium, trapezium left angle=70, trapezium right angle=110, minimum width=2cm, minimum height=1.5cm, text centered, draw=black, fill=white, text width=3cm, trapezium stretches=true]
\tikzstyle{line} = [draw=black!50, thick, -Latex]


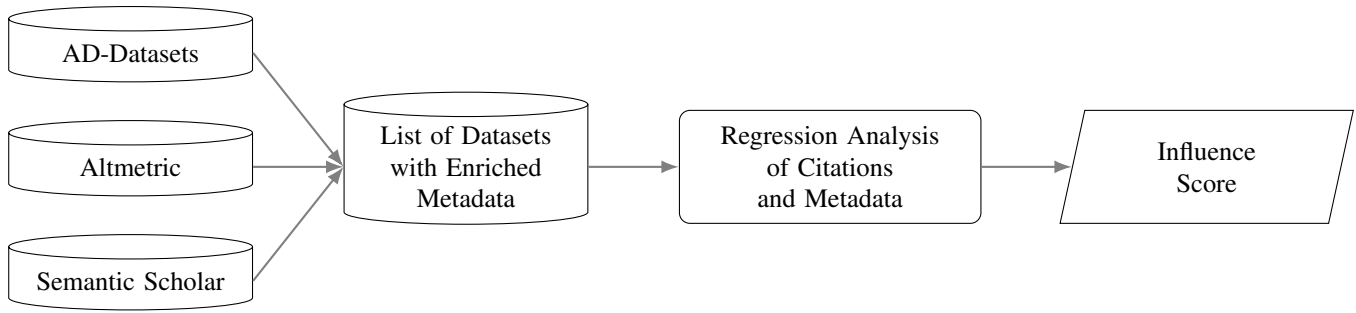
\begin{figure*}[h!]
    \noindent\resizebox{\textwidth}{!}{
        \begin{tikzpicture}[
                nodes={minimum height=3em, text width=7em, align=center},
                node distance = 5mm and 12mm,
                block/.style = {draw, rounded corners, fill=#1,
                        minimum height=3em, text width=8em, align=center},
                block/.default = white,
                every edge/.append style = {draw=black!50, thick, -Latex}
            ]
            \node[data] (A) {AD-Datasets};
            \node[data, below = of A] (B) {Altmetric};
            \node[data, below = of B] (C) {Semantic Scholar};
            \node[data, right = of B] (D) {List of Datasets with Enriched Metadata};
            \node[process, right = of D] (E) {Regression Analysis of Citations and Metadata};
            \node [io, right = of E] (F) {Influence\\Score};

            \draw[thick,anchor=west]   (A.east)  edge  (D.west);
            \draw[thick,anchor=west]   (B.east)  edge  (D.west);
            \draw[thick,anchor=west]   (C.east)  edge  (D.west);
            \draw[thick,anchor=west]   (D.east)  edge  (E.west);
            \draw[thick,anchor=west]   (E.east)  edge  (F.west);

        \end{tikzpicture}
    }
    \caption{Overview: First, we collect data from various sources and combine them to a single list of datasets. Based on this, we perform a regression analysis to determine which metadata correlate with future prediction counts. Based on the metadata, we compute our Influence Score (IS).}
    \label{fig:system_figure}
\end{figure*}
\section{Regression Analysis}
\label{sec:method}

Here, we first introduce our taxonomy of terms related to the assessment of datasets. Subsequently, we introduce our data sources. Based on these, we describe the regression analysis of citations and metadata. In Section~\ref{sec:evaluation}, we present the resulting Influence Score. Figure~\ref{fig:system_figure} gives an overview over this process.

\subsection{Taxonomy}

As became clear in Section~\ref{sec:related_work}, no common language for specific aspects in the domain has evolved yet. Thus, we introduce a taxonomy to clearly describe different aspects with respect to the development of a dataset or paper. As general terms for this, we utilize success, progress, performance, or potential. For concrete aspects, we establish the following terms, where each one can be applied to any single paper:

\textbf{Impact:} We use the number of citations to measure the scientific impact of a paper, which is common in Scientometrics, but not without criticism~\cite{bornmann_what_2008}.

\textbf{Attention:} The online reception, such as tweets and Wikipedia articles mentioning a paper, represents the attention by researchers and the public.

\textbf{Influence:} We refer to the resulting score of our proposed method, which combines a multitude of aspects, as the influence, or \ac{IS}, of a dataset. We deem this term appropriate for any method that goes beyond purely impact-based assessment.

\subsection{Data Sources and Selection}

We used three sources for our data: ad-datasets.com~\cite{addatasets_web}, the Semantic Scholar Academic Graph API~\cite{allen_institute_for_ai_semantic_nodate}, and altmetric.com~\cite{altmetriccom_altmetric_nodate}. Based on the DOI and arXiv-Id from ad-datasets, we automatically extracted the metadata of papers from Semantic Scholar and altmetric.com. Based on these papers, all of which describe datasets, we performed data exploration, regression, and the computation of the \ac{IS}.\\

\textbf{AD-Datasets:} This web tool offers an overview of over 200 data sets in \ac{ad} ~\cite{Bogdoll_Addatasets_2022_VEHITS}. It includes a detailed breakdown of most dataset entries by 20 different meta categories, provided by the authors and the research community. This way, relations between datasets, accompanying papers and further metadata are available. The underlying data is stored in the JSON format and can be accessed accordingly. In this work, we utilize the $n_{frames}$ and $n_{sensors}$ metadata, which indicate the size of datasets in different dimensions, which is a potential aspect of the relevance of a dataset.\\

\textbf{Altmetric:} We used the API by altmetric.com~\cite{altmetriccom_altmetric_nodate}, which provides insight into online attention and readership. These properties are provided by the following categories:\\
\textbf{Attention Score:} The $aas_{curr}$ aggregates different sources into a single score~\cite{williams_altmetric_2016}. It is a weighted count of different online sources. For example, the weight for a reference on Wikipedia is 3, while Twitter and Reddit mentions are both weighted with 0.25.
Unfortunately, the history of this score is only provided for the most recent year.\\
\textbf{Attention Score after three months:} The $aas_{3m}$ is the percentile of the papers' Attention Score three months after publication. The percentile is calculated in comparison to papers that have been released at a similar time.\\
\textbf{Readers:} The number of people $n_{readers}$ that have saved a paper in their reference management software. Reading a paper is less significant than citing it, but the number of readers might imply interest in a paper early on. The number of readers is provided individually for multiple reference management services, which we sum into a single count for online readers. Altmetric.com cannot verify the number of readers, thus, it is not included in the attention score. This is a relevant attribute, as it decouples the metrics. However, there are no historic data available.\\

\textbf{Semantic Scholar:} For every accompanying paper of a dataset, we pulled data from Semantic Scholar. Sometimes, multiple datasets are described in the same paper, which will lead to the same information for those datasets. We extracted the following nested data:
\begin{itemize}
    \item List of referenced papers, including for each a list of all citing papers and the year of citation.
    \item List of authors and their respective publications, including for each publication a list of citing papers and the year of citation.
    \item List of citing papers, including for each a list of citing papers and the year of citation.
\end{itemize}

Wherever possible, we collected associated timestamps, including the publication year $a_{pub}$ of each paper. The first two categories, while dynamic, are directly available. We use the citations of references as a measure of the impact of references. Having impactful references might indicate that a paper is covering popular topics within \ac{ad} or that the authors are knowledgeable in the field.

The performance of authors can be estimated by evaluating their paper count and how many citations they have received, which becomes only meaningful over time. As discussed earlier, not every citation means usage of a dataset. While it would have been interesting to take into account, in which section a paper has been cited, this data was not available for most papers. Based on the $n_{cit3}$ citations from the previous three years, citations of works that cited a dataset signify the value created by working with the dataset, which is why we included those.\\

A critical aspect of the collected data is that oftentimes, no historic information was available. Also, oftentimes, data was not available due to limitations, e.g., Altmetric is incompatible with DOIs from IEEE publications, which are common in the fields of robotics, autonomous driving, and machine learning. Similarly, Ravenscroft et al.~\cite{ravenscroft_measuring_2017} expressed concerns about Altmetric, as they were unable to find 40~\% of the papers they analyzed, all from the field of computer science.

\subsection{Data Aggregation}
To further utilize the raw data we collected, we aggregated some of it with the aim to assemble a finite list of features that describe a dataset. We aggregated some of our data sources using the concept of the h-index formula, as it is widely known, transparent, and easy to reproduce. In order to analyze smaller timespans, we deviated from the typical 5-year duration and calculated multiple 3-year indexes ourselves.

For authors, we applied the h3-index for each individual. We then aggregated the h-indices of all authors of a paper via the arithmetic mean in $aut_{\mu h3}$. Respectively, we applied the h-index formula to references and citations. For the references of a paper, the $ref_{h3}$ is calculated identically to the way it is utilized for authors. Just like an author has papers with citations, a paper has references with citations. A high h-index for references would signify that several of the referenced papers gained lots of attraction. We also applied the h3-index formula to the citations and their citations to get the h3-index of citations $cit_{h3}$, following Schubert et al.~\cite{schubert_using_2008}. The final list of all extracted and calculated features can be found in Table~\ref{tab:overview_feautres}.

\begin{table*}[t]
    \caption{Overview of metadata used for the regression analysis and the influence score.}
    \resizebox{\textwidth}{!}{
        \begin{tabular}{lllcccr}
            \toprule
            Feature        & Description                             & Availability                 & Standardized & Log(x+1)   & Influence Score & Source                                                         \\ \midrule
            $n_{frames}$   & Number of frames in the dataset         & At publication               & --   & -- & \checkmark      & AD-Datasets~\cite{addatasets_web}                              \\
            $n_{sensors}$  & Number of sensor types                  & At publication               & \checkmark   & \xmark     & \checkmark      & AD-Datasets~\cite{addatasets_web}                              \\
            $a_{pub}$            & Year of publication                     & At publication               & \checkmark   & \xmark     & \xmark          & Semantic Scholar~\cite{allen_institute_for_ai_semantic_nodate} \\
            $ref_{h3}$     & 3 year h-index of references            & At publication               & \checkmark   & \xmark     & \checkmark      & Semantic Scholar~\cite{allen_institute_for_ai_semantic_nodate} \\
            $aut_{\mu h3}$ & Mean 3 year h-index of authors papers   & At publication               & \checkmark   & \xmark     & \checkmark      & Semantic Scholar~\cite{allen_institute_for_ai_semantic_nodate} \\
            $n_{cit3}$     & Number of citations within past 3 years & Anytime after publication    & \checkmark   & \checkmark & \checkmark      & Semantic Scholar~\cite{allen_institute_for_ai_semantic_nodate} \\
            $cit_{h3}$     & 3 year h-index of citations             & $>$3 years after publication & --           & --         & \checkmark      & Semantic Scholar~\cite{allen_institute_for_ai_semantic_nodate} \\
            $aas_{curr}$   & Altmetric Attention Score               & Anytime after publication    & --           & --         & \checkmark      & Altmetric~\cite{altmetriccom_altmetric_nodate} \\      
            $aas_{3m}$ & Altmetric Attention Score at 3 mos             & After 3 months & \checkmark & \xmark & \xmark & Altmetric~\cite{altmetriccom_altmetric_nodate}    \\
            $n_{readers}$  & Number of readers                       & Anytime after publication    & --           & --         & \checkmark      & Altmetric~\cite{altmetriccom_altmetric_nodate}             \\ \bottomrule\\
        \end{tabular}
    }
    \label{tab:overview_feautres}
\end{table*}

\subsection{Cluster Analysis and Regression Setup}
\begin{figure}
    \includegraphics[width=\columnwidth]{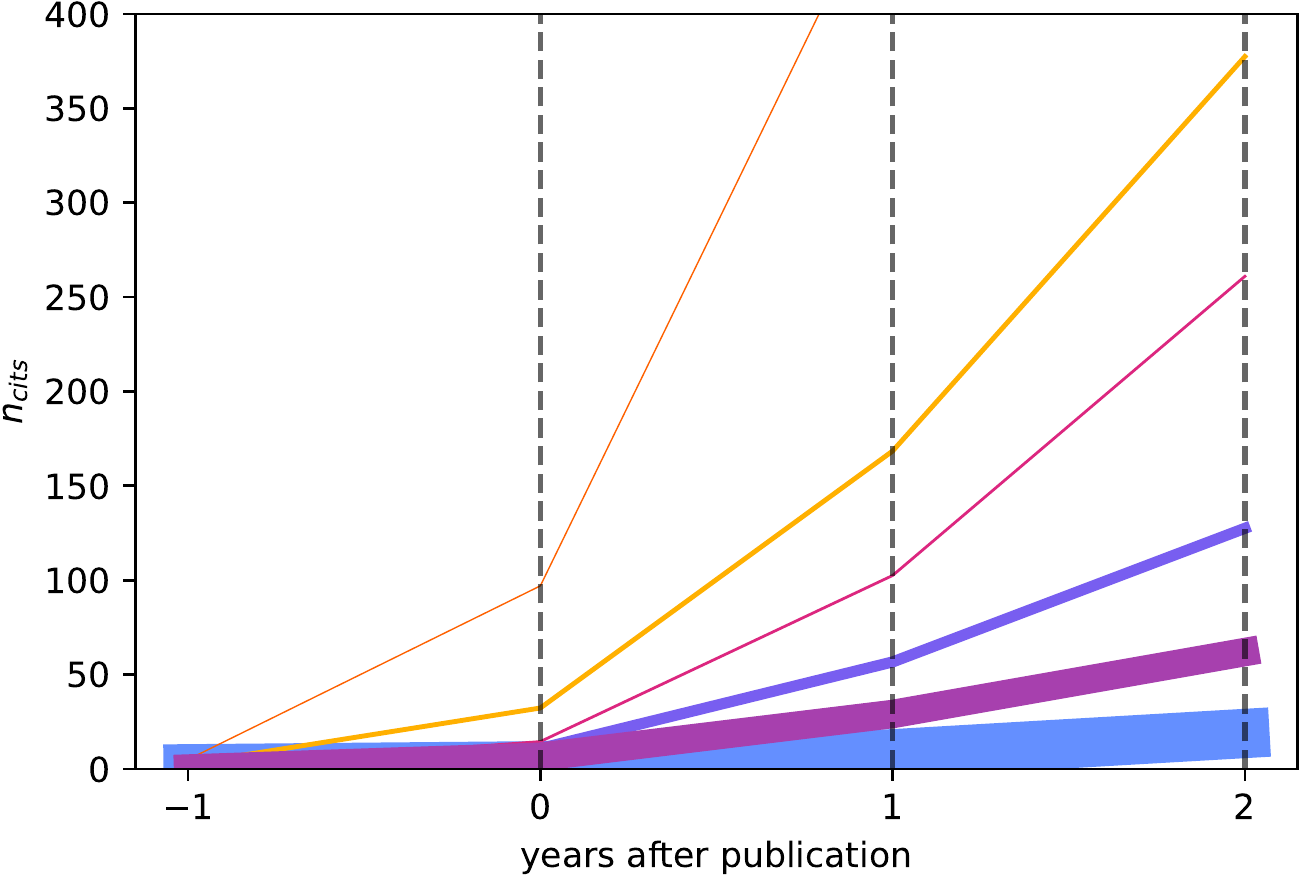}
    \centering
    \caption{Development of the number of citations for publication-clusters over a dynamic 3-year window. Papers are clustered based on similar performance. Semantic Scholar also tracks citations of pre-prints, which leads to citations prior to the publication date of the final work.}
    \label{fig:citations_clustering}
\end{figure}
We evaluated our computed features with respect to their ability to predict future citations. Therefor, we performed linear regression. For this, we first computed clusters of the datasets to determine a meaningful time horizon. Subsequently, we defined our regression setup.

\textbf{Cluster Analysis:}
To show that there are meaningful variations between clusters, we looked at the impact of papers for up to 2 years after publication in a journal or conference proceedings. As visualized in \figref{fig:citations_clustering}, clear clusters are visible, where line-thickness indicates cluster size. For k-means clustering, we used six clusters based on the elbow plot, which shows which additional cluster provides a non-marginal reduction of the total variation within clusters. The growth of citation counts behaves exponentially for the top performing works. A clear differentiation between all clusters becomes apparent already after one year, which we chose as the time horizon for the regression. This allowed us to include more recent papers, which would have been excluded otherwise due to their missing track record of citations.

\textbf{Regression Setup:}
As independent variables, we included the features $n_{sensors}$, $a_{pub}$, $ref_{h3}$, $aut_{\mu h3}$, $n_{cit3}$, and $aas_{3m}$, as shown in Table~\ref{tab:overview_feautres}, in order to estimate the citation count after one year. Preliminary data exploration suggested a curvilinear relationship between $aas_{3m}$ and the number of citations. Therefore, a quadratic term was added. All predictors were standardized by subtracting the mean and dividing by the standard deviation prior to the analysis. The feature $n_{cit3}$ was log(x+1)-transformed to ensure a normal distribution of the residuals, which are the error terms of the regression.

For the regression, we were able to utilize 111 datasets, as values for all included features were available, and they had been released at least one year prior. Residuals and collinearity, the ability to linearly predict one independent variable with other independent variables, were checked. The collinearity was quantified through the variance inflation factor of each regressor which all were lower than three. We performed the Breusch-Pagan and White test for heteroskedasticity, which is the inconsistency of the variance of residuals at different levels of the dependent variable. Both tests indicated that we do not have sufficient evidence for the presence of heteroskedasticity. Still, robust standard errors were used to ensure the standard errors are calculated correctly in the presence of heteroskedasticity which at worst leads to standard errors being estimated larger.

We chose not to include $n_{frames}$ for the regression because numerous of the datasets did not contain this meta-information. However, we examined a model in which the feature was included, which did not lead to new findings.

\subsection{Regression Analysis}
With the explained regression setup, we were now interested in finding statistically significant predictor variables for the citation count at the end of the year after publication.

The $aas_{3m}$ and $aas_{3m}^2$ were positively related to the number of citations and both relationships were significant at $<$0.0001. Both coefficients were positive. All other features were not significantly related to the number of citations. The results are reported in Table \ref{tab:regression_three_years}.

\begin{table}[ht]
    \caption{Regression for citations after one year. Regression coefficients and 95\% confidence interval are represented on the log scale.}
    \resizebox{\columnwidth}{!}{
        \begin{tabular}{lcccccc}
            \toprule
                           & coef    & std err & z      & P\textgreater{$\lvert$z$\rvert$} & {[}0.025 & 0.975{]} \\ \midrule
$ref_{h3}$                            & 0.0987 & 0.103   & 0.96  & 0.337              & -0.103   & 0.3      \\
$aut_{\mu h3}$                        & 0.071  & 0.086   & 0.824 & 0.41               & -0.098   & 0.24     \\
$a_{pub}$                                   & 0.0281 & 0.084   & 0.337 & 0.736              & -0.136   & 0.192    \\
$n_{sensors}$                         & 0.1383 & 0.108   & 1.287 & 0.198              & -0.072   & 0.349    \\
$aas_{3m}$          & 0.803  & 0.116   & 6.895 & \textbf{0}         & 0.575    & 1.031    \\
$aas_{3m}^2$ & 0.3375 & 0.072   & 4.72  & \textbf{0}         & 0.197    & 0.478    \\ \midrule
$intercept$                           & 2.6402 & 0.117   & 22.48 & 0                  & 2.41     & 2.87     \\ \bottomrule\\
        \end{tabular}
    }
    \label{tab:regression_three_years}
\end{table}

Since only one feature showed a relationship with the number of citations, we do not consider a stable prediction of citations possible with the available data. In order to still perform an early evaluation of datasets, in the following we present our Influence Score (IS).

\begin{figure*}[t!]
    \includegraphics[width=\textwidth]{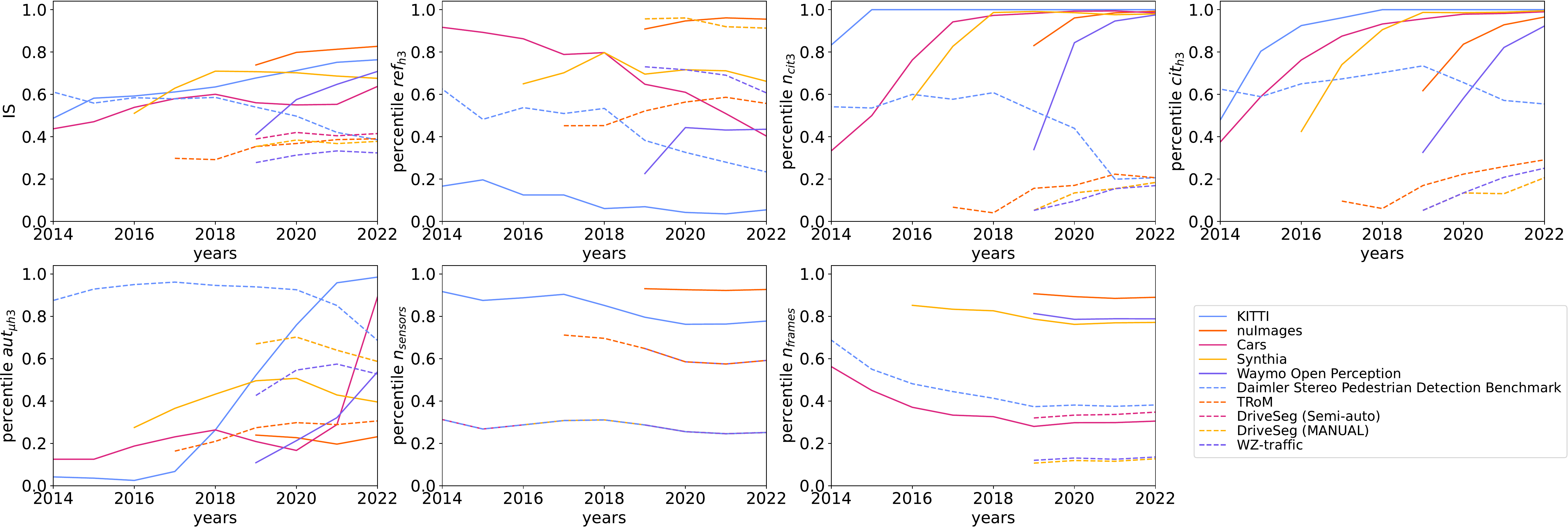}
    \centering
    \caption{Influence Score and individual features for different datasets. We show exemplary results for the five best and worst performing datasets of all time, measured by citations, with a latest release in 2019 for historical data. We also show six individual features of the \ac{IS}, where historical data was available.}
    \label{fig:influence_score}
\end{figure*}

\section{Influence Score}
\label{sec:evaluation}

We propose the Influence Score (IS), which includes a variety of features that are available early on. These are weighted dynamically in order to receive an indication of the relative performance of any given dataset at any given time. The calculation compares each dataset with all existing ones from the domain, so that relative differences and trends are immediately recognizable. 

Percentiles are used to allow relative scoring within the surrounding group of datasets. The datasets roughly follow a normal distribution in their \ac{IS} scores. As shown in Table~\ref{tab:overview_feautres}, we utilize eight different features for the \ac{IS}: $n_{frames}$, $n_{sensors}$, $ref_{h3}$, $aut_{\mu h3}$, $n_{cit3}$, $cit_{h3}$, $aas_{curr}$ and $n_{readers}$. This way, we consider more than just the citations, but do not exclude them: If early citations are already available, they become a meaningful part of the score, as the relation to other datasets of the peer group is relevant. This way, citation velocity is included. The \ac{IS} is defined as follows:

\begin{equation}
    \label{equation_influence_score}
    IS(paper) = 1/n * \sum_{i=0}^{n} percentile(feature_i) 
\end{equation}

where:
\begin{conditions}
 i     &  Feature Index \\
 n     &  Number of available features \\
\end{conditions}

Only features, which are available, are dynamically included in the \ac{IS}. As we used percentiles of each feature to facilitate the understanding of the features, common issues are mitigated. E.g., typical feature values change over time: For example, with the growth of \ac{ad}, the $n_{cit}$ value of a paper today is likely higher than a decade ago, which becomes clearly visible in Figure~\ref{fig:growth_of_publications}. Furthermore, commonly observed values for features might differ between different fields. This helps people who are not familiar with \ac{ad} or the features to easily assess if the score a dataset achieved is high or low.

\subsection{Qualitative Demonstration}

To showcase the \ac{IS}, we compare exemplary the development of the five most and least cited papers with a latest publication in 2019, by their \ac{IS} and visualize the results in \figref{fig:influence_score}. It becomes clearly visible, that the two groups are easily distinguishable by their \ac{IS}, but also that differences within the groups are visible. 

The individual features show different pictures: For $ref_{h3}$, also papers with only a few citations can have meaningful references in their works. $n_{cit3}$ and $cit_{h3}$ only confirm what was known by our data selection, as we selected the datasets by citation count. $aut_{\mu h3}$ shows, how successful datasets can also boost personal careers, as some authors became professors and remained active in their field. $n_{sensors}$ and $n_{frames}$ show rather static results, with a trend towards larger datasets being more successful.

\subsection{Quantitative Demonstration}

\begin{figure}[ht]
    \includegraphics[width=\columnwidth]{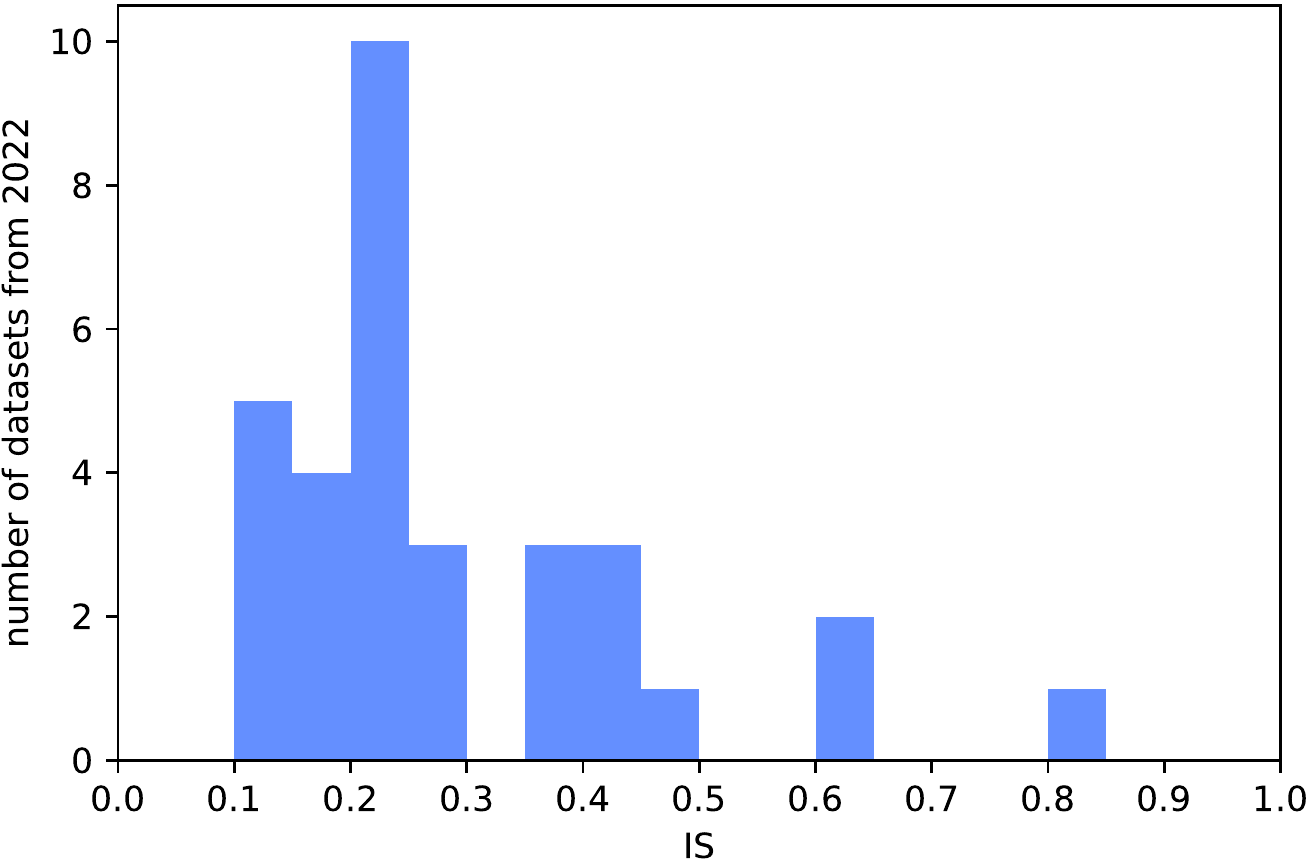}
    \centering
    \caption{Distribution of the Influence Score (IS) of all datasets from 2022.}
    \label{fig:hist_is_2022}
\end{figure}

In order to show the quantitative performance of the \ac{IS}, we showcase all datasets released in 2022 in a detailed overview in Table~\ref{tab:influenceScoreThreeYears2022}. Such a pre-filtering process is useful in order to explore novel datasets. Here, it becomes clear that the IS captures a wide variety of different aspects of a dataset. Of particular interest is the fact that even low-performing datasets can lead in certain features. Thus, if a researcher is interested in certain aspects of a dataset, they can simply focus on the features they are interested in and omit the others, which enables less-known datasets to be discovered and used. Figure~\ref{fig:hist_is_2022} shows an overview of the IS distributions.

\begin{table*}[]
\caption{Influence Score and features for datasets released in 2022. Sorted by IS, top 3 features bold.}
\resizebox{\textwidth}{!}{%
\begin{tabular}{@{}lccccccccc@{}}
\toprule
                                                   & $\ac{IS}$ & $n_{cit3}$  & $cit_{h3}$ & $ref_{h3}$ & $aut_{\mu h3}$ & $n_{frames}$ & $n_{sensors}$ & $aas_{curr}$ & $n_{readers}$ \\ \midrule
Waymo Block-NeRF~\cite{tancikBlockNeRFScalableLarge2022}                                   & 0.82           & \textbf{0.7}  & \textbf{0.53}  & \textbf{0.95} & 0.87    & --     & --      & \textbf{1.0}       & \textbf{0.89}    \\
SHIFT~\cite{sunSHIFTSyntheticDriving2022}                                              & 0.62           & 0.25 & 0.2   & \textbf{0.99} & \textbf{0.88}    & \textbf{0.94}   & \textbf{0.77}    & 0.65      & \textbf{0.42}    \\
Street Hazards~\cite{hendrycksScalingOutofDistributionDetection2022}                                     & 0.62           & \textbf{0.67} & \textbf{0.58}  & 0.83 & \textbf{0.92}    & 0.16   & 0.25    & \textbf{0.73}      & \textbf{0.45}    \\
KITTI-360-APS~\cite{mohanAmodalPanopticSegmentation2022}                                      & 0.48           & 0.23 & 0.06  & 0.68 & 0.67    & 0.56   & 0.25    & \textbf{0.97}      & 0.21    \\
ScribbleKITTI~\cite{unalScribbleSupervisedLiDARSemantic2022}                                      & 0.45           & 0.2  & 0.15  & 0.83 & \textbf{0.97}    & 0.32   & 0.25    & 0.63      & 0.02    \\
BDD100K-APS~\cite{mohanAmodalPanopticSegmentation2022}                                        & 0.42           & 0.23 & 0.06  & 0.68 & 0.67    & 0.1    & 0.25    & \textbf{0.97}      & 0.21    \\
Ithaca365~\cite{diaz-ruizIthaca365DatasetDriving2022}                                          & 0.4            & 0.15 & 0.15  & 0.68 & 0.22    & \textbf{0.82}   & \textbf{0.58}    & 0.53      & 0.26    \\
CODA~\cite{liCODARealWorldRoad2022}                                              & 0.39           & 0.2  & 0.15  & 0.75 & 0.36    & --     & 0.25    & 0.53      & 0.33    \\
Rope3D~\cite{yeRope3DTheRoadsidePerception2022}                                             & 0.37           & 0.15 & 0.15  & 0.83 & 0.42    & 0.53   & \textbf{0.58}    & 0.25      & 0.25    \\
Comma2k19 LD~\cite{satoDrivingOrientedMetricLane2022}                                       & 0.37           & 0.12 & 0.15  & 0.71 & 0.56    & --     & --      & 0.64      & 0.02    \\
RoadSaW~\cite{cordesRoadSaWLargeScaleDataset2022}                                            & 0.29           & 0.09 & 0.06  & 0.27 & 0.19    & \textbf{0.83}   & 0.25    & --        & --      \\
K-Radar~\cite{paekKRadar4DRadar2022}                                            & 0.26           & 0.09 & 0.06  & 0.47 & 0.11    & 0.44   & \textbf{0.93}    & 0.4       & 0.26    \\
CARLA-WildLife~\cite{maagTwoVideoData2022}                               & 0.26           & 0.03 & 0.06  & \textbf{0.92} & 0.12    & --     & 0.25    & 0.34      & 0.08    \\
AugKITTI~\cite{panUnderstandingChallengesWhen2022}                                           & 0.25           & 0.03 & 0.06  & 0.88 & 0.16    & --     & --      & 0.25      & 0.1     \\
WildDash 2~\cite{zendelUnifyingPanopticSegmentation2022}                                         & 0.24           & 0.17 & 0.2   & 0.52 & 0.08    & --     & 0.25    & --        & --      \\
MONA~\cite{gressenbuchMONAMunichMotion2022} & 0.23           & 0.03 & 0.06  & 0.36 & 0.48    & --     & 0.25    & --        & --      \\
Street Obstacle Sequences~\cite{maagTwoVideoData2022}                    & 0.23           & 0.03 & 0.06  & \textbf{0.92} & 0.12    & 0.07   & 0.25    & 0.34      & 0.08    \\
HDBD~\cite{qiuIncorporatingGazeBehavior2022}                                               & 0.22           & 0.03 & 0.06  & 0.16 & 0.64    & --     & \textbf{0.58}    & --        & --      \\
GLARE~\cite{grayGLAREDatasetTraffic2022}                                              & 0.22           & 0.03 & 0.06  & 0.47 & 0.29    & --     & --      & 0.42      & 0.06    \\
Boreas~\cite{burnettBoreasMultiSeasonAutonomous2022}                                             & 0.22           & \textbf{0.29} & \textbf{0.25}  & 0.14 & 0.2     & --     & --      & --        & --      \\
Autonomous Platform Inertial~\cite{shurinAutonomousPlatformsInertial2022}                       & 0.21           & 0.2  & \textbf{0.25}  & 0.36 & 0.03    & --     & --      & --        & --      \\
aiMotive~\cite{matuszkaAiMotiveDatasetMultimodal2022}                                           & 0.2            & 0.03 & 0.06  & 0.33 & 0.04    & 0.41   & \textbf{0.93}    & 0.44      & 0.13    \\
CarlaScenes~\cite{kloukiniotisCarlaScenesSyntheticDataset2022}                                        & 0.2            & 0.03 & 0.06  & 0.52 & 0.2     & --     & \textbf{0.77}    & --        & --      \\
LUMPI~\cite{buschLUMPILeibnizUniversity2022}                                              & 0.19           & 0.09 & 0.06  & 0.02 & 0.07    & 0.74   & \textbf{0.58}    & --        & --      \\
A9~\cite{cressA9DatasetMultiSensorInfrastructureBased2022}                                                 & 0.19           & 0.25 & \textbf{0.25}  & 0.14 & 0.1     & --     & \textbf{0.58}    & 0.29      & 0.12    \\
Amodal Cityscapes~\cite{breitensteinAmodalCityscapesNew2022}                                  & 0.19           & 0.09 & 0.06  & 0.27 & 0.43    & 0.11   & 0.25    & 0.29      & 0.08    \\
R-U-MAAD~\cite{wiedererBenchmarkUnsupervisedAnomaly2022}                                           & 0.16           & 0.03 & 0.06  & 0.33 & 0.35    & --     & 0.25    & 0.15      & 0.06    \\
TJ4DRadSet~\cite{zhengTJ4DRadSet4DRadar2022}                                         & 0.15           & 0.12 & 0.15  & 0.14 & 0.05    & --     & --      & 0.42      & 0.02    \\
OpenMPD~\cite{zhangOpenMPDOpenMultimodal2022}                                            & 0.14           & 0.18 & 0.15  & 0.02 & 0.07    & 0.27   & \textbf{0.58}    & --        & --      \\
I see you~\cite{quispeSeeYouVehiclePedestrian2022}                                          & 0.12           & 0.03 & 0.06  & 0.09 & 0.01    & --     & --      & 0.44      & 0.08    \\
SceNDD~\cite{prabuSceNDDScenariobasedNaturalistic2022}                                             & 0.11           & 0.09 & 0.06  & 0.16 & 0.19    & --     & --      & 0.15      & 0.02    \\
exiD~\cite{moersExiDDatasetRealWorld2022}                                               & 0.11           & 0.12 & 0.06  & 0.02 & 0.24    & --     & 0.25    & --        & --      \\ \bottomrule\\
\end{tabular}%
}
\label{tab:influenceScoreThreeYears2022}
\end{table*}

\section{Conclusion}
\label{sec:conclusion}
In this paper, we addressed the lack of knowledge with respect to the scientific impact, attention, and influence of datasets in robotics. Our focus was on an early assessment of datasets, given a flood of new datasets published every year. We analyzed impact measured by citations and evaluated relations of metadata and features which we extracted from multiple online sources. Our regression analysis showed no strong relation between future citations and our selected features. Subsequently, we presented our developed Influence Score (IS). This score utilizes a set of eight features to assess any dataset also early on. This is based on an analysis within the peer group of all datasets, which allows for the early detection of relative trends.

Our work contributes to a better understanding of datasets, which enables researchers to find and assess published datasets in the domain of autonomous driving without the need of waiting for a track record of citations.\\

\textbf{Limitations and Outlook}:
For our work, we evaluated the paper accompanying the dataset assuming that the paper is a good representation of the dataset. When measuring scientific impact through citations, we think this holds because the paper is actually the cited scientific work. However, not every citation might be meaningful, positive, or indicate the usage of a dataset. Ideally, large language models could evaluate if a dataset is actually used, if cited. Khan et al.~\cite{khan_measuring_2021}, who analyzed datasets in biodiversity, suggested that the correlation between the number of downloads and citations signifies that these two measures are comparable representations of impact. However, in the domain of \ac{ad}, download numbers are typically not available, but this might change. As some datasets are presented in the same paper, a further decoupling of accompanying papers and the respective datasets would be helpful. We found, that the quality and availability of metadata in \ac{ad} provided by the creators of datasets varies strongly. Thus, standards should be established~\cite{google_datasets}. While we focussed on dataset and paper specific features for this work, we are also interested in the venue or journal of publication, which can be considered as an additional feature in future work.
\section{Acknowledgment}
\label{sec:acknowledgment}
This work results partly from the KIGLIS project supported by the German Federal Ministry of Education and Research (BMBF), grant number 16KIS1231. We want to thank both Altmetric and Semantic Scholar, who have provided us with the necessary API accesses for this work.


{\small
  \bibliographystyle{IEEEtran}
  \bibliography{references.bib}
}

\end{document}